\documentclass[preprint,prd,tightenlines,nofootinbib]{revtex4}
\usepackage{graphicx}
\font\symbolfont=cmsy10 at 10 pt
\def\texttilde{{\symbolfont \char'030}}
\def \MSbar {\vbox{\hrule\kern 1pt\hbox{\rm MS}}}
\begin{document}

\title{Next-to-leading order QCD calculations with parton showers II:
soft singularities}

\author{ Davison E.\ Soper}
\affiliation{Institute of Theoretical Science, 
University of Oregon, Eugene, OR 97403 USA}
\date{19 February 2004}

\begin{abstract}
Programs that calculate observables in quantum chromodynamics at
next-to-leading order typically generate events that consist of
partons rather than hadrons -- and just a few partons at that. These
programs would be much more useful if the few partons were turned into
parton showers, which could be given to one of the Monte Carlo event
generators to produce hadron showers. In a previous paper, we have seen
how to generate parton showers related to the final state collinear
singularities of the perturbative calculation for the example of $e^+ +
e^- \to 3\ {\rm jets}$. This paper discusses the treatment of the soft
singularities.
\end{abstract}

\maketitle

\section{Introduction}

This is the second of two papers concerning how to modify a
next-to-leading order (NLO) calculation in quantum chromodynamics by
adding parton showers in such a way that when the generated events are
used to calculate an infrared safe observable, the observable is
calculated correctly at next-to-leading order. In the companion paper
\cite{I}, Kr\"amer and the present author explain the aspects of this
problem that relate to the divergences that appear in perturbation theory
when two massless partons become collinear. This part of the method is
quite simple and easy to understand. The treatment of the divergences that
appear when a gluon becomes soft is a bit more technical and will be
given in this paper.

We find that it it is conceptually simplest to match showers to a NLO
calculation if the NLO calculation is performed in the Coulomb gauge.
That is because the collinear divergences of the theory in a physical
gauge like the Coulomb gauge are isolated in the cut and virtual
self-energy diagrams -- just the diagrams that you would draw to
illustrate a parton splitting to two partons. Accordingly, the present
research program was initiated in Ref.~\cite{KSCoulomb}, which shows how
to do NLO calculations in the Coulomb gauge. I should point out that,
although there is a certain conceptual simplicity that emerges through
the use of a physical gauge, one can use any gauge, as the recent work of
Frixione, Nason, and Webber on the same subject demonstrates
\cite{FrixioneWebberI, FrixioneWebberII}.

I begin with a brief restatement of the problem. Consider the calculation
of an experimental observable that describes a hard scattering process
involving strongly interacting particles and suppose that the chosen
observable is infrared safe, which makes a perturbative calculation for
the observable sensible. The simplest sort of calculation for this
purpose consists of calculating the hard process cross section at the
lowest order in $\alpha_s$, call it $\alpha_s^B$, at which it occurs.
However, one often performs a next-to-leading order calculation,
including terms proportional to $\alpha_s^B$ and $\alpha_s^{B+1}$. Then
the estimated error from yet higher order terms (which are usually
unknown) is typically smaller than with a leading order calculation.

The NLO calculations are typically in the form of computer programs
that act as Monte Carlo generators of partonic events. They produce
simulated events with final states $f$ consisting of a few partons with
specified momenta. For example, for three-jet production in
electron-positron annihilation, there are three or four partons in the
final state. Each event comes with a weight $w_n$. Then the predicted
value for an observable described by a measurement function
${\cal S}(f)$ is
\begin{equation}
{\cal I} = 
\lim_{N\to \infty}{ 1 \over N}\sum_{n=1}^N
w_n\,{\cal S}(f_{\!n}).
\label{average}
\end{equation}

A serious weakness of most current NLO Monte Carlo generators is that they
produce simulated final states that are not close to physical final
states. This contrasts sharply with the highly successful leading order
Monte Carlo event generators such as Pythia, Herwig, and Ariadne
\cite{Pythia,Herwig,Ariadne}, which do generate realistic final states.
What could be done to generate realistic final states in an NLO
calculation? One might think of using the three and four parton final
states in the NLO partonic generators to generate a parton shower from
each outgoing parton. This gives a final state with lots of partons. Then
one could use one of the leading order Monte Carlo event
generators to turn the many partons into hadrons. There should not be a
serious problem with the hadronization stage, since this is modelled as a
long distance process that leaves infrared safe observables largely
unchanged. The problem lies with the parton showers. Here a high energy
parton splits into two daughter partons, which each split into two more
partons. As this process continues, the virtualities of successive pairs
of daughter partons gets smaller and smaller, representing splittings
that happen at larger and larger distance scales. The late splittings
leave infrared safe observables largely unchanged. However, the first
splittings in a parton shower can involve large virtualities. They
represent a mixture of long distance and short distance physics. Thus one
must be careful that the showering does not reproduce some piece of short
distance physics that was already included in the NLO calculation. To be
a little more precise, if one expands the prediction of the calculation
including showers in a power series in $\alpha_s(Q)$, where $Q$ is the
hard process scale, in the form
\begin{equation}
{\cal I} =  C_0\, \alpha_s^B + C_1\, \alpha_s^{B+1}
+ C_3\, \alpha_s^{B+2} +\cdots,
\end{equation}
then our goal is to insure that the coefficients $C_0$ and $C_1$ are
exactly the same in the calculation with showers as in the purely
perturbative NLO calculation.

The algorithms that we propose apply to showers from final state massless
partons. They are illustrated using the process $e^+ + e^- \to 3\ {\rm
{jets}}$ in quantum chromodynamics.\footnote{I discuss $e^+ + e^- \to 3\
{\rm {jets}}$ but not $e^+ + e^- \to 2\ {\rm {jets}}$.  Thus the
present calculation is to be applied with a measurement function that
give zero for events that are close to the two-jet configuration.}  While
our research program was underway, Frixione and Webber succeeded in
matching showers to a NLO calculation for a problem with massless partons
in the initial state (but no observed colored partons in the final state)
\cite{FrixioneWebberI}. Subsequently, Frixione, Webber, and Nason have
extended this method to a problem with massive colored partons in the
final state \cite{FrixioneWebberII}. Their method is similar to the method
presented in Ref.~\cite{I}, or more precisely to a variant of that method
in which the first level of parton splitting, based on Coulomb gauge
splitting functions in Ref.~\cite{I}, is instead based on the splitting
functions of Herwig. The soft gluon effects treated in this paper are not
part of the algorithms of  \cite{FrixioneWebberI, FrixioneWebberII}.
Instead, these references employ what amounts to a soft gluon 
cutoff.\footnote{The parameter $C_1$
above is then modified by an amount proportional to a positive power of
the cutoff parameter $\beta$, so as long as
$\beta$ is chosen small enough, the cutoff effects 
in \cite{FrixioneWebberI,FrixioneWebberII} are negligible.}
There has been other work on this problem \cite{otherwork}. That
of Collins \cite{JCC} is particularly instructive.

\section{The problem of the soft singularities}

In this section, I describe the particular problem left over from
Ref.~\cite{I} that remains to be addressed in this paper, recalling just
enough of the argument of Ref.~\cite{I} to set up the issue.

\begin{figure}
\includegraphics[width = 8 cm]{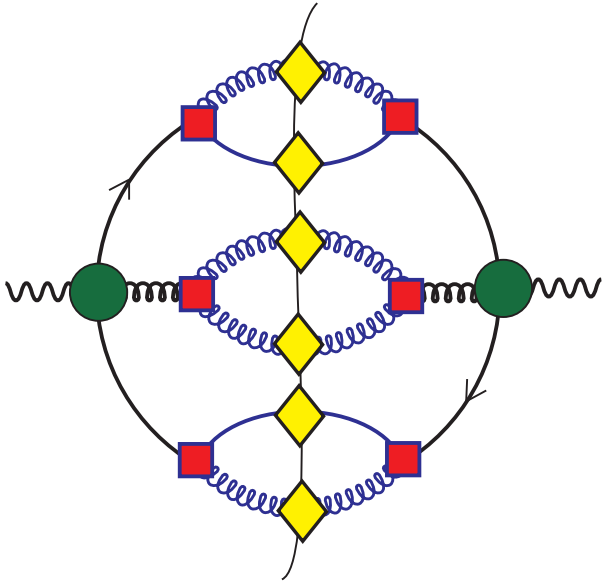}
\medskip
\caption{Primary parton splitting from Ref.~\protect\cite{I}. The filled
circles represent graphs for the Born amplitude and complex conjugate
amplitude. Each of the partons emerging from the Born amplitude splits
into two partons with a vertex, represented by the squares, that
includes a Sudakov suppression factor. Each of the six daughter partons
undergoes further, secondary, splittings and enters the final state as a
complete shower. The secondary splittings are represented by the
diamonds.}
\label{fig:shower1}
\end{figure}

Consider one of the Born graphs for $e^+ + e^- \to 3\ {\rm jets}$. Three
partons emerge into the final state. The idea presented in Ref.~\cite{I}
is to let each of these partons split into a pair of partons, as
illustrated in Fig.~\ref{fig:shower1}. Two features of the splitting are
important. First, when the virtuality $\bar q^2$ of the pair is small, the
splitting probability is proportional to $(P(x)/\bar q^2) d\bar q^2 dx$
where $x$ is the share of the momentum carried by one of the daughter
partons and $P(x)$ is the appropriate Altarelli-Parisi splitting
function. Second, the collinear singularity at $\bar q^2 \to 0$ is damped
by a Sudakov factor with the behavior $\exp(-\alpha_s\,  c\log^2(\bar
q^2))$ for $\bar q^2 \to 0$. Thus each of the three partons makes a jet,
but in the limit of small $\alpha_s$, each jet is usually very narrow and
appears in an infrared safe measurement like a single massless parton.
There is a qualifying adverb ``usually'' here. A fraction $\alpha_s$ of
the time, one of the splittings has a substantial virtuality and we get a
four-jet final state. Thus there is an order $\alpha_s^{B+1}$ effect in
which some probability is removed from the three-jet final state and
given to a four-jet final state. In a purely NLO calculation, this effect
is included in the $\alpha_s^{B+1}$ graphs. Thus, to keep the calculation
correct to NLO, we subtract these probabilities from the $\alpha_s^{B+1}$
perturbative graphs.

The subtractions are required for NLO consistency of the calculation, but
they have another important effect. Since the splitting functions have
just the structure found in the order $\alpha_s^{B+1}$ graphs in the
collinear limit, the subtractions cancel the collinear singularities (and
the collinear $\times$ soft  singularities) of the order $\alpha_s^{B+1}$
graphs. The real and virtual singularities are separately cancelled,
whereas without the subtractions the real and virtual singularities are
left to cancel against each other (which, of course, they do\footnote{
We employ a calculation \cite{beowulfPRL, beowulfPRD, beowulfrho} in which
the integrals over the three-momenta in virtual loops are performed
numerically. Then this cancellation happens automatically point by point
in the integration. In other methods, the virtual integrals are performed
entirely analytically. In this case, the cancellation mechanism is less
simple. }). If there were no other singularities, we could let the three
and four parton final states of the order $\alpha_s^{B+1}$ graphs develop
into parton showers, which would affect the result of the calculation
only at order $\alpha_s^{B+2}$. The calculation would be numerically
stable because the integrals for the three and four parton states would be
separately finite.

But, of course, there are other singularities, namely the singularities
from emission of soft gluons at wide angles from the three hard partons,
as well as the corresponding singularities in the $\alpha_s^{B+1}$
virtual graphs. These singularities cancel between the order
$\alpha_s^{B+1}$ graphs with three and four final state partons, but we
cannot let these two kinds of final states develop separately into 
showers without losing this cancellation and thus encountering numerical
instability.

This is not just a technical problem to be somehow sidestepped. The
emission of a soft gluon at order $\alpha_s^{B+1}$ is a real physical
effect. It is an effect that does not fit a picture of separate evolution
of the three jets because the soft gluon is sensitive to the whole final
state. That is, the probability to emit a soft gluon does not break up
into the sum of three independent probabilities, one for emission from
each jet.

\begin{figure}
\includegraphics[width = 8 cm]{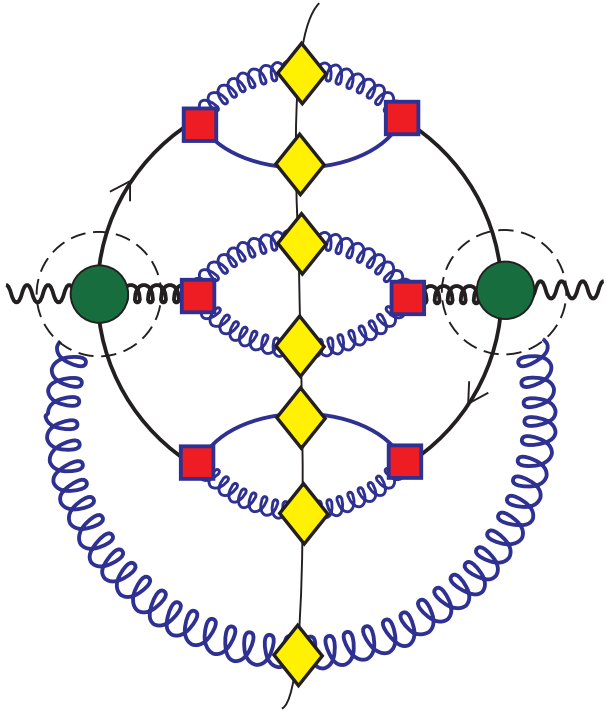}
\medskip
\caption{Primary parton splitting with soft gluon radiation added. The
graphical symbols of Fig.~\ref{fig:shower1} are also used here. The extra
gluon coming from the circles on the right and left represents the soft
gluon radiated from the three jets. As with the other partons, the soft
gluon undergoes secondary splittings and generates a shower, represented
by the diamonds.}
\label{fig:shower2}
\end{figure}

There is a natural way to deal with this. We can consider the three jets
to form an antenna that radiates a soft gluon. The radiation probability
is proportional to $(f(\theta,\phi)/E) dE d\Omega$, where $E$ is the
energy of the gluon, $\theta,\phi$ are its angles, and $f(\theta,\phi)$
is a function that is determined by the directions of the three jets. We
can account for the normalization of the radiation probability provided by
the virtual graphs by providing a suppression factor with the behavior
$\exp(-\alpha_s\,  c\ |\log(E)|)$ for $E \to 0$. This is depicted in
Fig.~\ref{fig:shower2}. Thus a soft gluon is always emitted, but when
$\alpha_s$ is small the gluon is usually too soft to matter to an
infrared safe measurement. However, a fraction
$\alpha_s$ of the time the gluon can have a substantial energy and we get
a four-jet final state. Thus there is an order $\alpha_s^{B+1}$ effect in
which some probability is removed from the three-jet final state and
given to a four-jet final state. In the purely NLO calculation, this
effect is included in the $\alpha_s^{B+1}$ graphs. Thus, to keep the
calculation correct to NLO, we subtract these probabilities from the
$\alpha_s^{B+1}$ perturbative graphs.\footnote{As we shall see, some of
the adjustment is to be made in the wide angle part of the splitting
functions used to generate the splittings of the three hard partons.} Of
course, the required subtraction terms will be just the soft singularity
subtractions that we needed to stabilize the $\alpha_s^{B+1}$
contributions if we add showers separately to $\alpha_s^{B+1}$ graphs
with three and four-jet final states.

Does the necessity to add soft radiation from a three-jet antenna mean
that the standard parton shower Monte Carlo programs such as Herwig,
Pythia, and Ariadne need to be rewritten in order to be compatible with
an NLO calculation? Not at all. We do not need soft radiation with its
correct angular dependence at every stage of a parton shower. All that we
need is to radiate one soft gluon from the hard event with its correct
angular dependence. This radiation, along with the initial splittings of
the three hard partons, can be considered to be part of the NLO
calculation, so that the Monte Carlo programs do not need any
modification at all. I will add more comments on this subject in the
conclusions section of this paper.  

\section{Soft gluon emission}

In Ref.~\cite{I}, we saw that one can turn ``${\it Born}\times[ 1 + {\it
real} - |{\it virtual}|]$'' into ``${\it Born}\times {\it
real}\,\exp(-|{\it virtual}|)$'' with respect to self-energy graphs,
which contain the collinear and ${\rm collinear}\times{\rm soft}$
singularities in the Coulomb gauge. In this section, we begin a detailed
description of the singularities generated by the emission of soft
gluons at wide angles to the outgoing partons. Up to now in our
description, these singularities cancel between real and virtual emissions
in the form ``${\it Born}\times[ 1 + {\it real} - |{\it virtual}|]$.'' In
the following sections, we turn this into the form ``${\it Born}\times
{\it real}\,\exp(-|{\it virtual}|)$'' together with some left over
nonsingular terms.

Consider a cut Born graph in which the final state has a quark
(momentum $q_1$), a gluon (momentum $q_2$) and an antiquark (momentum
$q_3$). These partons will make collinear showers, but we ignore that for
now. What concerns us here is that these outgoing partons act as an
antenna that radiates soft gluons. Let a soft gluon with momentum $l$ be
emitted by parton $i$ and, in the complex conjugate amplitude, absorbed
by parton $j$.

We need to specify the kinematics in a definite way. We choose to route
the momenta for the $\{i,j\}$ term so that the momenta $\{\vec
p^{\,(ij)}_1\!, \vec p^{\,(ij)}_2\!, \vec p^{\,(ij)}_3\}$ carried by,
respectively, the quark, the gluon, and the antiquark in the final state
are given by
\begin{equation}
\vec p^{\,(ij)}_k = \vec q_k - {\textstyle{1\over 2}}
(\delta_{ki} + \delta_{kj})\vec l .
\label{pkijdef}
\end{equation}
Thus for a graph with $i = j$, the momentum $k$ before the soft gluon
emission is $\vec k = \vec q$ and afterwards it is $\vec p = \vec q -
\vec l$. For a graph with $i \ne j$, the momentum $k$ before the soft
gluon emission is $\vec k = \vec q + {\textstyle{1\over 2}}\, \vec l$ and
afterwards it is $\vec p = \vec q - {\textstyle{1\over 2}}\, \vec l$.

\subsection{The soft gluon approximation}

We now specify the approximation to be applied for the soft gluon
emission \cite{GrammerYennie}. Let a parton with momentum $k$ emit a soft
gluon with momentum $l$ and enter the final state with momentum $p$. Then
$l^\mu = k^\mu - p^\mu$. The momenta $p$ and $k$ are approximately in the
direction of a light-like vector $u$. We take $u = (1,\hat q)$ where
$\hat q = \vec q/|\vec q\,|$, and $\vec q$ is
${\textstyle{1\over 2}}\,(\vec k + \vec p)$ in the case of an $i \ne j$
graph or $\vec k$ in the case of an $i = j$ graph, as specified in
Eq.~(\ref{pkijdef}).

The approximation is based on the fact that the soft gluon sees a current
$J^\mu$ that is approximately proportional to $u^\mu$. Thus if we denote
the polarization vector of the gluon by $\epsilon^\mu$, we can replace
\begin{equation}
J\cdot \epsilon \to J \cdot l\ { u\cdot \epsilon \over u\cdot l}.
\label{softapprox}
\end{equation}
This replacement becomes exact in the limit in which $J^\mu$
approaches $\lambda u^\mu$ for some constant $\lambda$. It is important
the the momentum $l$ not point along $u$. That is, the gluon momentum
should be small so that both $k$ and $p$ and thus $J$ are approximately
proportional to $u$, but the gluon momentum should not be collinear with
the momentum of the emitting parton.

For the graph in which the soft gluon goes from parton $i$ to parton $j$,
the factors ${ u\cdot \epsilon / u\cdot l}$, together with appropriate
color matrices, give a factor (in the Coulomb gauge)
\begin{equation}
C_{ij}\,
{ u_i^\mu\,u_j^\nu\,
D_{\mu\nu}(l)
\over 
(u_i^\alpha l_\alpha)
(u_j^\beta  l_\beta)
}
=
{C_{ij}\over \vec l^2}\,
{
(\hat q_i\cdot\hat q_j)-
(\hat l\cdot\hat q_i)(\hat l\cdot\hat q_j)
\over 
(1 - \hat l\cdot \hat q_i)\
(1 - \hat l\cdot \hat q_j)
}.
\label{soft}
\end{equation}
Here $D_{\mu\nu}(l)$ is the Coulomb gauge numerator factor
\begin{equation}
D^{\mu\nu}(l)
=
-g^{\mu\nu} 
+\frac{1}{\vec l^{\,2}}\left[
-l^\mu \tilde l^\nu
-\tilde l^\mu  l^\nu
+l^\mu  l^\nu
\right],
\end{equation}
where $\tilde l$ is $(0,l^1,l^2,l^3)$. The constants $C_{ij}$ are the
color factors for this process,
\begin{eqnarray}
&&C_{12} = C_{21} =  C_{23} = C_{32} = -{ N_C \over 2},
\nonumber\\
&&C_{13} = C_{31} = { 1 \over 2 N_C},
\nonumber\\
&&C_{11} = C_{33} = C_F,
\nonumber\\
&&C_{22} = C_A.
\label{colorfactors}
\end{eqnarray}

What about the factor $J\cdot l$? In the case of emission from a quark,
we have (since $p^2 = 0$)
\begin{equation}
\rlap{/}p\,
(\rlap{/}k - \rlap{/}p)\,
{ \rlap{/}k \over k^2}
= \rlap{/}p,
\label{softquark}
\end{equation}
so we simply include a factor $\rlap{/}p$ for the final state quark. The
final state phase space associated with the original graph has a factor
$1/(2|\vec p\,|)$. We retain this factor without change.

For
emission from an antiquark we have [noting that the momenta $p$ and $q$
are directed against the fermion number flow and that we should extract a
factor $(-1)$ to associate with the color factor for the graph]
\begin{equation}
(-1)\times
{(- \rlap{/}k) \over k^2}\,
(\rlap{/}k - \rlap{/}p)\,
(-\rlap{/}p)
= -\rlap{/}p,
\label{softqbar}
\end{equation}
so we simply include a factor $-\rlap{/}p$ for the final state antiquark.
The final state phase space associated with the original graph has a
factor $1/(2|\vec p\,|)$. We retain this factor without change.

The case of emission from a gluon is a bit more complicated. We denote
the tensor associated with the vertex function by
\begin{equation}
V^{\alpha\beta\gamma}(k_A,k_B,k_C)
= g^{\alpha\beta}(k_A^\gamma - k_B^\gamma)
+{\rm cyclic\ permutations}.
\end{equation}
Then the three gluon vertex function with momenta $\{k_A,k_B,k_C\}$
directed into the vertex is 
$
(-ig) (-if_{abc})V^{\alpha\beta\gamma}(k_A,k_B,k_C).
$
The vertex dotted into $(k-p)$ times the adjoining propagator and
the propagator numerator for the final state gluon, with the $-ig$ and
the color matrix $-if_{abc}$ removed, is
\begin{eqnarray}
\lefteqn{
D(p)_{\alpha}^{\mu}\,
\left[
V^{\alpha\beta}_\gamma(-p,k,p-k)\,(k^\gamma - p^\gamma)
\right]\,
{ D(k)^{\nu}_{\beta} \over k^2}\
}
\hskip 2 cm
&&
\nonumber\\ 
&=&
-D(p)_{\alpha}^{\mu} 
\left[
(k^2 g^{\alpha\beta} - k^\alpha k^\beta)
-
(p^2 g^{\alpha\beta} - p^\alpha p^\beta)
\right]
{ D(k)^{\nu}_{\beta} \over k^2}
\nonumber\\
&=&
-D(p)_{\alpha}^{\mu} 
(k^2 g^{\alpha\beta} - k^\alpha k^\beta)
{ D(k)^{\nu}_{\beta} \over k^2} 
\nonumber\\
&=&
D(p)_{\alpha}^{\mu} 
(g^{\alpha\nu} + { 1 \over \vec k^2}\ \tilde k^\alpha k^\nu )
\nonumber\\
&=&
D(p)_{\alpha}^{\mu} 
(g^{\alpha\nu} 
+ { 1 \over \vec k^2}\ (\tilde p^\alpha + \tilde l^\alpha) k^\nu )
\nonumber\\
&=&
D(p)_{\alpha}^{\mu} 
(g^{\alpha\nu} 
+ { 1 \over \vec k^2}\  \tilde l^\alpha k^\nu ) 
\nonumber\\
&\to& 
D(p)^{\mu\nu}.
\label{softgluon}
\end{eqnarray}
Here in the second to last step we have noted that $D(p)_{\alpha}^{\mu}
\tilde p^\alpha$ = 0.  Then in the last step we drop the term proportional
to $l$ so as to simplify our soft gluon approximation. Thus the net
result of the approximation is to include, in addition to the factor ${
u\cdot \epsilon / u\cdot l}$, a factor $D(p)^{\mu\nu}$ for the final
state gluon. The final state phase space associated with the original graph has a
factor $1/(2|\vec p\,|)$. We retain this factor without change.

It is important that the factors in Eqs.~(\ref{softquark}),
(\ref{softqbar}) and (\ref{softgluon}) revert to the factors in the Born
diagram when $\vec l \to 0$.

\subsection{The added terms}

We can now apply these ideas. We write the contribution to the cross
section from a given cut Born level graph as
\begin{eqnarray}
{\cal I}[{\rm Born}] &=&
\int\! {d\vec q_1} 
\int\! {d\vec q_2} 
\int\! {d\vec q_3}\ 
\delta\!\left(\sum \vec q_i\right)\
G_3(\vec q_1,\vec q_2, \vec q_3)
\nonumber\\
&&\times
f_3(\vec q_1,\vec q_2, \vec q_3)\
{\cal S}_3(\vec q_1,\vec q_2, \vec q_3).
\label{softbornfull}
\end{eqnarray}
Here ${\cal S}_3$ is the final state measurement function [see Eq.~(4) of
Ref.~\cite{I}] and $f_3$ contains the factors from the Feynman rules
corresponding to the three final state particles, which, suppressing the
Dirac and Lorentz indices, are
\begin{equation}
f_3(\vec q_1,\vec q_2, \vec q_3)
= \frac{\rlap{/}q_1}{2|\vec q_1|}\
\frac{D(q_2)}{2|\vec q_2|}\
\frac{-\rlap{/}q_3}{2|\vec q_3|}.
\end{equation}
The function $G_3$ contains everything else, including an amputated
Green function times a hermitian conjugate amputated Green function. 
With this notation, we write the terms to be added for real soft gluon
emission as
\begin{eqnarray}
{\cal I}[{\rm soft,real}] &=&
\int\! {d\vec q_1} 
\int\! {d\vec q_2} 
\int\! {d\vec q_3}\ 
\delta\!\left(\sum \vec q_i\right)\
G_3(\vec q_1,\vec q_2, \vec q_3)
\nonumber\\
&&\times
\frac{g_s^2}{(2\pi)^{3}}
\int\! \frac{d\vec l}{2|\vec l\,|}\
\frac{1}{\vec l^{\,2}}\
\theta( \vec l^2 < M_{\rm soft}^2)
\sum_{ij}
F_{ij}(\hat l;\hat q_1,\hat q_2, \hat q_3)\,
\nonumber\\
&&\times
f_3(\vec p^{\,(ij)}_1\!,\vec p^{\,(ij)}_2\!,\vec p^{\,(ij)}_3)\
{\cal S}_4(\vec p^{\,(ij)}_1\!,\vec p^{\,(ij)}_2\!,\vec p^{\,(ij)}_3\!,
\vec l\ ),
\label{softrealfull}
\end{eqnarray}
where $\vec p^{\,(ij)}_k$ are defined in Eq.~(\ref{pkijdef}) and ${\cal
S}_4$ is the measurement function for four final state particles.

The factor $\theta( \vec l^2 < M_{\rm soft}^2)$ is inserted
to suppress contributions from momenta $\vec l$ that are outside the
range of validity of the soft gluon approximation. We take
\begin{equation}
M_{\rm soft} \equiv \lambda_{\rm soft} \sqrt{s_0}\,(1-t_0),
\end{equation}
where $\sqrt{s_0}$ and $t_0$ are, respectively, the c.m.~energy and the
thrust of the final state with parton momenta $\{\vec q_1,\vec q_2, \vec
q_3\}$ and where we take the parameter $\lambda_{\rm soft}$ to be 1/3. 

The functions $F_{ij}$ are, for $i \ne j$,
\begin{equation}
F_{ij} = C_{ij}\, 
{ \hat q_i\cdot\hat q_j - (\hat l \cdot \hat q_i)(\hat l \cdot \hat q_j)
\over (1 - \hat l\cdot \hat q_i)(1 - \hat l\cdot \hat q_j)},
\end{equation}
as specified in Eq.~(\ref{soft}). For $i = j$ we begin with a factor
\begin{equation}
C_{ii}\,{1 - (\hat l \cdot \hat q_i)^2 \over (1 - \hat l \cdot \hat
q_i)^2} = C_{ii}\,{1 + \hat l \cdot \hat q_i \over 1 - \hat l \cdot \hat
q_i}.
\end{equation}
We use Eq.~(\ref{colorfactors}) to break $C_{ii}$ up into
\begin{equation}
C_{ii} = - \sum_{a = 1\atop a \ne i}^3 C_{ia}.
\end{equation}
Then in the term proportional to $C_{ia}$, we impose the condition that
the angle between $\vec l$ and $\vec q_i$ should be greater than the angle
between $\vec q_a$ and $\vec q_i$. That is, our soft gluon for $i=j$
should be soft but not collinear, and we impose a cut to enforce that
condition. Thus we take
\begin{equation}
F_{ii} = { 1 + \hat l \cdot \hat q_i \over 1 - \hat l \cdot \hat q_i}\
\biggl[
(-1) \sum_{a = 1\atop a \ne i}^3 C_{ia}\,
\theta(\hat l \cdot \hat q_i <  \hat q_a \cdot \hat q_i)
\biggr].
\end{equation}

It will be useful to have at hand an abbreviated notation based on the
definitions
\begin{equation}
R_0(\vec q_1,\vec q_2, \vec q_3)
=
G_3(\vec q_1,\vec q_2, \vec q_3)\
f_3(\vec q_1,\vec q_2, \vec q_3)\
{\cal S}_3(\vec q_1,\vec q_2, \vec q_3)
\end{equation}
and
\begin{equation}
R_{ij}(\vec l;\vec q_1,\vec q_2, \vec q_3)
=
G_3(\vec q_1,\vec q_2, \vec q_3)\
f_3(\vec p^{\,(ij)}_1\!,\vec p^{\,(ij)}_2\!,\vec p^{\,(ij)}_3)\
{\cal S}_4(\vec p^{\,(ij)}_1\!,\vec p^{\,(ij)}_2\!,\vec p^{\,(ij)}_3\!,
\vec l\ ).
\end{equation}
With this notation, the Born and soft-real contributions are
\begin{equation}
{\cal I}[{\rm Born}] =
\int\! {d\vec q_1} 
\int\! {d\vec q_2} 
\int\! {d\vec q_3}\ 
\delta\!\left(\sum \vec q_i\right)\
R_0(\vec q_1,\vec q_2, \vec q_3)
\label{softborn}
\end{equation}
and
\begin{eqnarray}
{\cal I}[{\rm soft,real}] &=&
\int\! {d\vec q_1} 
\int\! {d\vec q_2} 
\int\! {d\vec q_3}\ 
\delta\!\left(\sum \vec q_i\right)
\int_0^{M_{\rm soft}} { d|\vec l\,| \over |\vec l\,|}\,
\int { d^2\hat l \over 4 \pi}\
\nonumber\\
&&\times
\sum_{ij}
{\alpha_s \over \pi}\,
F_{ij}(\hat l;\hat q_1,\hat q_2, \hat q_3)\,
R_{ij}(\vec l;\vec q_1,\vec q_2, \vec q_3).
\label{softreal}
\end{eqnarray}
The indicated integration over $\hat l$ is an integration over the unit
sphere.

We define a corresponding virtual soft gluon contribution that will
cancel the real soft gluon contribution for small $\vec l$ as long as
the measurement function is infrared safe, so that $R_{ij} \to R_0$ for
$\vec l \to 0$: 
\begin{eqnarray}
{\cal I}[{\rm soft,virtual}] &=&
-
\int\! {d\vec q_1} 
\int\! {d\vec q_2} 
\int\! {d\vec q_3}\ 
\delta\!\left(\sum \vec q_i\right)
\int_0^{M_{\rm soft}} { d|\vec l\,| \over |\vec l\,|}\,
\int { d^2\hat l \over 4 \pi}\
\nonumber\\
&&\times
\sum_{ij}
{\alpha_s \over \pi}\,
F_{ij}(\hat l;\hat q_1,\hat q_2, \hat q_3)\,
R_0(\vec q_1,\vec q_2, \vec q_3).
\label{softvirtual}
\end{eqnarray}

Conveniently, the integral over the angles of $\vec l$ 
in ${\cal I}[{\rm soft,virtual}]$ is simple:
\begin{eqnarray}
\int \frac{d^2 \hat l}{4\pi}
\sum_{ij}
F_{ij}(\hat l;\hat q_1,\hat q_2, \hat q_3)
&=&
\sum_{i=1}^2 \sum_{j = i+1}^3\
(-C_{ij})(1-\hat q_i\cdot\hat q_j)
\nonumber\\
&=&
\frac{N_C}{2}[
(1-\hat q_1\cdot\hat q_2) + (1-\hat q_3\cdot\hat q_2)
]
-\frac{1}{2 N_C}
(1-\hat q_1\cdot\hat q_3) .\ \ \ \
\label{angularint}
\end{eqnarray}
The integral is largest when $\hat q_1\cdot\hat q_3 = 1$, which implies
$\hat q_1\cdot\hat q_2 = \hat q_3\cdot\hat q_2 = -1$. Then the integral
is $2N_C = 2C_A$. The integral is smallest when $\hat q_1\cdot\hat q_2 =
1$, or else $\hat q_3\cdot\hat q_2 = 1$. Then the integral is $N_C -
1/N_C = 2 C_F$.

We will, in due course, see how to exponentiate the added contributions
${\cal I}[{\rm soft,real}] + {\cal I}[{\rm soft,virtual}]$. First,
however, we must subtract these terms from ${\cal I}$ so that the cross
section remains unchanged.

\section{Adjusting the collinear terms}

The terms in Eqs.~(\ref{softreal}) with $i=j$ represent soft gluon
approximations to the cut self-energy diagrams for outgoing partons $i$.
Thus we should subtract them from the exact cut self-energy diagrams.
Consider, for example, $i = j = 1$, corresponding to a self-energy
insertion on the outgoing quark propagator. This contribution has the form
given in Eq.~(7) of Ref.~\cite{I},
\begin{equation}
{\cal I}[{\rm real}] =
\int\! {d\vec q\over  2|\vec q\,|}\ {\rm Tr}\left\{ 
\int_0^\infty\! {d\bar q^2 \over \bar q^2}
\int_0^1\!dx
\int_{-\pi}^\pi\! {d\phi\over 2\pi}\
\frac{\alpha_s}{2\pi}{\cal M}_{g/q}(\bar q^2,x,\phi)\,
R(\bar q^2,x,\phi) 
\right\},
\label{quarkjet}
\end{equation}
where $q = q_1$ and the notation, including the variables $(\bar q^2,
x,\phi)$, is explained in Sec.~III of Ref.~\cite{I}. We can translate the
$\{1,1\}$ term in Eq.~(\ref{softreal}) into the notation used in this
equation by using the change of variable
\begin{equation}
\int \frac{d\vec l}{2|\vec q - \vec l|\,2|\vec l\,|}\cdots
= 
\frac{1}{8|\vec q|}
\int_0^\infty\! {d\bar q^2}
\int_0^1\!dx
\int_{-\pi}^\pi\! 
{d\phi}\ { 1 \over 1 + \Delta}\cdots,
\end{equation}
where
\begin{equation}
1+\Delta = \sqrt{1 + \bar q^2/\vec q^{\,2}}.
\end{equation}
Then
\begin{eqnarray}
{\cal I}_{11}[{\rm soft, real}]&=&
\int\! {d\vec q\over  2|\vec q\,|}\ {\rm Tr}\Biggl\{ 
\int_0^\infty\! {d\bar q^2 \over \bar q^2}
\int_0^1\!dx
\int_{-\pi}^\pi\! 
{d\phi\over 2\pi}\ { 1 \over 1 + \Delta}\,
\frac{\bar q^2}{\vec l^{\,2}}
{ \alpha_s \over 4\pi}\
\theta( \vec l^{\,2} <  M_{\rm soft}^2)
\nonumber\\
&&\times
\frac{1 + \hat l\cdot \hat q}{1 - \hat l\cdot \hat q}\
\biggl[
(-1) \sum_{a = 1\atop a \ne 1}^3 C_{1a}\,
\theta\!\left(
\hat l\cdot \hat q
< \hat q_a \cdot \hat q\right)
\biggr]\
(\rlap{/}q - \rlap{/}\,l)\,
R(\bar q^2,x,\phi) 
\Biggr\}.
\label{I11}
\end{eqnarray}
Now we can use
\begin{eqnarray}
 \vec l^{\,2}&=&\frac{\vec q^{\,2}}{4}\,(\Delta + 2x)^2
\nonumber\\
\bar q^2 &=&\vec q^{\,2} \Delta (2 + \Delta)
\nonumber\\
\hat l \cdot \hat q &=& \frac{2x - \Delta + 2x\Delta}{\Delta + 2 x}
\end{eqnarray}
to rewrite this as
\begin{eqnarray}
{\cal I}_{11}[{\rm soft, real}]&=&
\int\! {d\vec q\over  2|\vec q\,|}\ {\rm Tr}\Biggl\{ 
\int_0^\infty\! {d\bar q^2 \over \bar q^2}
\int_0^1\!dx
\int_{-\pi}^\pi\! 
{d\phi\over 2\pi}\ { 1 \over 1 + \Delta}\,
\theta(\Delta + 2x < 2M_{\rm soft}/|\vec q\,|)
\nonumber\\
&&\times
\biggl[
(-1) \sum_{a = 1\atop a \ne 1}^3 C_{1a}\,
\theta\!\left(
{ 2x - \Delta + 2 x \Delta \over \Delta + 2 x}
< \hat q_a \cdot \hat q\right)
\biggr]
\nonumber\\
&&\times
{ \alpha_s \over 2\pi}\
{ 8 x \over (\Delta + 2 x)^2}\
{ (1 + \Delta/2)^2 \over 1-x}\
(\rlap{/}q - \rlap{/}\,l)\,
R(\bar q^2,x,\phi) 
\Biggr\}.
\label{I11bis}
\end{eqnarray}

Thus we can subtract ${\cal I}_{11}[{\rm soft, real}]$ by replacing
${\cal M}_{g/q}(\bar q^2,x,\phi)$ by
\begin{equation}
{\cal M}_{g/q}(\bar q^2,x,\phi) 
- (\rlap{/}q - \rlap{/}\,l)\,
{\cal P}^{{\rm soft}}_1(\bar q^2,x),
\end{equation}
where
\begin{eqnarray}
{\cal P}^{{\rm soft}}_i(\bar q^2,x) &=&
\theta(\Delta + 2x < 2M_{\rm soft}/|\vec q\,|)\,
\biggl[
(-1) \sum_{a = 1\atop a \ne i}^3 C_{ia}\,
\theta\!\left(
{ 2x - \Delta + 2 x \Delta \over \Delta + 2 x}
< \hat q_a \cdot \hat q\right)
\biggr]\
\nonumber\\
&& \times
{ 8 x \over 1 - x}\
{ 1 \over (\Delta + 2 x)^2}\
{ (1 + \Delta/2)^2 \over 1 + \Delta}.
\label{calPsoft}
\end{eqnarray}
The effect of this is to remove the part of ${\cal M}$ that
is singular when $|\vec l\,|\to 0$ at a fixed angle greater than the
angle between the quark and the other two outgoing partons. Note that
$|\vec l\,|\to 0$ at a fixed angle corresponds to $\Delta \to 0$ and $x
\to 0$ with $\Delta/x$ fixed.

The term in Eq.~(\ref{softreal}) for emission from the antiquark line,
${\cal I}_{33}[{\rm soft, real}]$ is given by Eq.~(\ref{I11bis}) with  
$(\rlap{/}q - \rlap{/}\,l)$ replaced by $(-\rlap{/}q + \rlap{/}\,l)$ and
the index 1 replaced by 3. Thus we can subtract ${\cal I}_{33}[{\rm soft,
real}]$ by replacing ${\cal M}_{g/\bar q}(\bar q^2,x,\phi)$ in the
analogue of Eq.~(\ref{quarkjet}) by
\begin{equation}
{\cal M}_{g/\bar q}(\bar q^2,x,\phi) 
- (-\rlap{/}q + \rlap{/}\,l)\,
{\cal P}^{{\rm soft}}_3(\bar q^2,x),
\end{equation}
with the function ${\cal P}^{{\rm soft}}_3(\bar q^2,x)$ given in
Eq.~(\ref{calPsoft}) with $i = 3$.

Finally, the term in Eq.~(\ref{softreal}) for emission from the gluon
line, ${\cal I}_{22}[{\rm soft, real}]$ is given by 
\begin{eqnarray}
{\cal I}_{22}[{\rm soft, real}]&=&
\int\! {d\vec q\over  2|\vec q\,|}\ 
\int_0^\infty\! {d\bar q^2 \over \bar q^2}
\int_0^1\!dx
\int_{-\pi}^\pi\! 
{d\phi\over 2\pi}\
\nonumber\\
&&\!\times
{ \alpha_s \over 2\pi}\,
{\cal P}^{\rm soft}_2(\bar q^2,x)
D^{\mu\nu}(q - l)
R_{\mu\nu}(\bar q^2,x,\phi) ,
\label{I22}
\end{eqnarray}
with the function ${\cal P}^{{\rm soft}}_2(\bar q^2,x)$ given in
Eq.~(\ref{calPsoft}) with $i = 2$.  Thus we can subtract ${\cal
I}_{22}[{\rm soft, real}]$ by replacing ${\cal M}^{\mu\nu}_{g/g}(\bar
q^2,x,\phi)$ in the analogue of Eq.~(\ref{quarkjet}) by
\begin{equation}
{\cal M}^{\mu\nu}_{g/g}(\bar q^2,x,\phi) 
-  D^{\mu\nu}(q - l)\,
{\cal P}^{{\rm soft}}_2(\bar q^2,x).
\end{equation}
(There is also a $g \to q\bar q$ splitting function ${\cal
M}^{\mu\nu}_{q/g}$, but there is no soft singularity for $g \to q\bar q$
so this function does not get modified.)

The terms in Eq.~(\ref{softvirtual}) with $i=j$ represent soft gluon
approximations to the virtual self-energy diagrams for outgoing partons
$i$. Thus we should subtract them from the exact virtual self-energy
diagrams. Consider the virtual self-energy insertion on the outgoing
quark propagator. This contribution has the form given in Eq.~(12) of
Ref.~\cite{I},
\begin{equation}
{\cal I}[{\rm virtual}] =
\int\! {d\vec q\over  2|\vec q\,|}\ {\rm Tr}\left\{
- 
\int_0^\infty {d\bar q^2 \over \bar q^2}
\int_0^1\! dx
\int_{-\pi}^\pi\! {d\phi\over 2\pi}\ 
{ \alpha_s \over 2\pi}\,
{\cal P}_{\!g/q}(\bar q^2,x)\,\rlap{/}q\, R_0 
\right\}.
\label{quarkjetvirt}
\end{equation}
The functions ${\cal P}_{\!a/b}$ are given in the Appendix of
Ref.~\cite{I}. When translated into the notation used in this equation,
the $\{1,1\}$ term in Eq.~(\ref{softvirtual}) is
\begin{equation}
{\cal I}_{11}[{\rm soft, virtual}] =
\int\! {d\vec q\over  2|\vec q\,|}\ {\rm Tr}\left\{
- 
\int_0^\infty {d\bar q^2 \over \bar q^2}
\int_0^1\! dx
\int_{-\pi}^\pi\! {d\phi\over 2\pi}\ 
{ \alpha_s \over 2\pi}\,
{\cal P}^{\rm soft}_1(\bar q^2,x)\,\rlap{/}q\, R_0 
\right\},
\label{I11virt}
\end{equation}
where ${\cal P}^{\rm soft}_1$ is defined in Eq.~(\ref{calPsoft}).
Thus we can subtract ${\cal I}_{11}[{\rm soft, virtual}]$ by replacing
${\cal P}_{\!g/q}(\bar q^2,x)$ by
\begin{equation}
{\cal P}_{\!g/q}(\bar q^2,x) 
- 
{\cal P}^{{\rm soft}}_1(\bar q^2,x).
\end{equation}
Similarly, we can subtract ${\cal I}_{33}[{\rm soft, virtual}]$ by
replacing ${\cal P}_{\!g/\bar q}(\bar q^2,x) = {\cal P}_{\!g/q}(\bar
q^2,x)$ in the analogue of Eq.~(\ref{quarkjetvirt}) by
\begin{equation}
{\cal P}_{\!g/q}(\bar q^2,x) 
- 
{\cal P}^{{\rm soft}}_3(\bar q^2,x)
\end{equation}
and we can subtract ${\cal I}_{22}[{\rm soft, virtual}]$ by
replacing ${\cal P}_{\!g/g}(\bar q^2,x)$ in the analogue of
Eq.~(\ref{quarkjetvirt}) by
\begin{equation}
{\cal P}_{\!g/g}(\bar q^2,x) 
- 
{\cal P}^{{\rm soft}}_2(\bar q^2,x).
\end{equation}

After the collinear splitting terms have been adjusted to remove their
wide angle soft parts, we turn ``${\it Born}\times[ 1 + {\it real} - |{\it
virtual}|]$'' for the collinear splittings into ``${\it Born}\times {\it
real}\,\exp(-|{\it virtual}|)$'' as explained in Ref.~\cite{I}.
The only difference is that the meaning of ``${\it real}$'' and ``${\it
virtual}$'' here have been slightly changed.

\section{Adjusting the NLO graphs}

It remains to subtract the terms ${\cal I}_{ij}[{\rm soft, real}]$ and
${\cal I}_{ij}[{\rm soft, virtual}]$ for $i \ne j$ from ${\cal I}$.
Recall that we include in the calculation the cut order $\alpha_s^{B+1}$
graphs that do not have a cut self-energy subdiagram or a virtual
self-energy subdiagram with the immediately adjacent propagator cut. (We
also include diagrams with cut self-energy subdiagrams, but with a
theta function that requires the virtuality in the self-energy subdiagram
to be large.)  The included graphs have soft gluon divergences, which
cancel between real and virtual soft gluon emissions. We will subtract the
approximate soft gluon contributions from these exact contributions, so
that the soft gluon divergences are cancelled separately for the real and
virtual contributions.

\subsection{Real soft gluons}

This is simple for the real emission terms, ${\cal I}[{\rm soft,
real}]$. Consider a cut order $\alpha_s^{B+1}$ graph with four parton
lines cut. The cut lines are a quark and an antiquark plus either two
gluons or another quark and antiquark. In the case of  a quark and an
antiquark plus another quark and antiquark, we do nothing. In the case of
a quark and an antiquark plus two gluons, we designate one of the gluons
as potentially soft. Then this cut graph corresponds to the contribution
from one of the possible Born graphs to one of the terms in ${\cal
I}[{\rm soft,real}]$, which has two gluons in the final state, one of
them ``hard'' and one soft. We simply subtract this contribution to
${\cal I}[{\rm soft, real}]$ from the cut graph in question.  Now we
designate the other final state gluon as potentially soft, find the
corresponding contribution to ${\cal I}[{\rm soft, real}]$ and subtract
it.

This procedure has two effects. First, we use up all of the terms in 
${\cal I}[{\rm soft, real}]$ that we needed to subtract from ${\cal I}$.
Second, the cut next-to-leading order graph with these counter-terms has
only integrable soft-gluon singularities since the singularities of
the counterterms match the singularities of the cut order
$\alpha_s^{B+1}$ graph when either of the gluons become soft.
Furthermore, the cancelling terms have the same final states: we are
cancelling graphs with four final state partons against counterterms with
the same four final state partons.

\subsection{Virtual soft gluons}

Subtracting the virtual terms, ${\cal I}[{\rm soft, virtual}]$ requires a
bit more analysis. We want to subtract these terms from the cut
next-to-leading order graphs with three partons in the final state -- a
quark, a gluon, and an antiquark. Such a cut graph has a virtual loop
either to the right of the final state cut or to the left. We want
the contributions from $-{\cal I}[{\rm soft, virtual}]$ to cancel the
soft gluon singularities from these virtual loops. However, the terms in 
${\cal I}[{\rm soft, virtual}]$, as written, do not have the right
structure to effect the cancellation. Thus we need to rearrange the terms.

In the $\{i,j\}$ term in ${\cal I}[{\rm soft, virtual}]$ we have the
function
\begin{equation}
- F_{ij}(\hat l) = - C_{ij}\, 
{ \hat q_i\cdot\hat q_j - (\hat l \cdot \hat q_i)(\hat l \cdot \hat q_j)
\over (1 - \hat l\cdot \hat q_i)(1 - \hat l\cdot \hat q_j)}.
\end{equation}
In the $\{j,i\}$ term we choose to redefine the integration variable so
that $\hat l \to - \hat l$. Then we have a contribution with
\begin{equation}
- F_{ij}(-\hat l) = - C_{ij}\, 
{ \hat q_i\cdot\hat q_j - (\hat l \cdot \hat q_i)(\hat l \cdot \hat q_j)
\over (1 + \hat l\cdot \hat q_i)(1 + \hat l\cdot \hat q_j)}.
\end{equation}
We use the identity
\begin{equation}
- F_{ij}(\hat l) - F_{ij}(-\hat l)
= F_{ij}^{L}(\hat l) + F_{ij}^{R}(\hat l),
\end{equation}
where
\begin{eqnarray}
F_{ij}^{L}(\hat l) &=&
- C_{ij}\
{ \hat q_i\cdot\hat q_j - (\hat l \cdot \hat q_i)(\hat l \cdot \hat q_j)
 \over (\hat q_j - \hat q_i)\cdot \hat l + i\epsilon}
\left(
{ 1 \over 1 + \hat q_i\cdot \hat l} + 
{ 1 \over 1 - \hat q_j\cdot \hat l}
\right)
\nonumber\\
&&+ { 2 C_{ij} \over (\hat q_j - \hat q_i)\cdot \hat l + i\epsilon}
\label{FijL}
\end{eqnarray}
and
\begin{eqnarray}
F_{ij}^{R}(\hat l) &=&
- C_{ij}\
{ \hat q_i\cdot\hat q_j - (\hat l \cdot \hat q_i)(\hat l \cdot \hat q_j)
 \over (\hat q_i - \hat q_j)\cdot \hat l - i\epsilon}
\left(
{ 1 \over 1 - \hat q_i\cdot \hat l} + 
{ 1 \over 1 + \hat q_j\cdot \hat l}
\right)
\nonumber\\
&&+ { 2 C_{ij} \over (\hat q_i - \hat q_j)\cdot \hat l - i\epsilon}.
\label{FijR}
\end{eqnarray}

The function $F_{ij}^{L}$ gives the singular factor that
arises from the exchange of a soft virtual gluon in a loop graph to the
left of the final state cut. Similarly, $F_{ij}^{R}$ gives the soft
gluon singularity for a loop graph to the right of the final state cut.
The essential part of a demonstration of this results from
considering the integral
\begin{equation}
I = 2|\vec l|^3 \int{ dl^0 \over 2\pi}\ 
{iC_{ij}\,u_i^\mu u_j^\nu D_{\mu\nu}(l) \over 
(u_i\cdot l + i\epsilon)
(-u_j\cdot l + i\epsilon)
(l^2 + i\epsilon)}.
\end{equation}
Here we have used the notation of Eq.~(\ref{soft}) for the emission and
absorption of a real gluon and have applied the same approximations. Upon
performing the integration, we find
\begin{equation}
I = F_{ij}^{L}(\hat l).
\end{equation}
In $F_{ij}^{L}(\hat l)$, the two parts of the first term represent
respectively the process in which parton $j$ emits a (transversely
polarized) soft gluon that is absorbed by parton $i$ and the process in
which parton $i$ emits a soft gluon that is absorbed by parton $j$. The
last term is the phase produced by the Coulomb force between the outgoing
partons. 

With these functions at hand, we consider a cut next-to-leading order
graph with three parton lines cut. Then there is a virtual loop. In the
virtual loop there may be one or two gluons that connect two lines that
go to the final state. Designate one of these gluons as potentially
soft. This graph then corresponds to one of the contributions to 
${\cal I}[{\rm soft,virtual}]$. We subtract a term of the form
\begin{eqnarray}
{\cal I}_{ij}^J[{\rm soft,virtual}] &=&
\int\! {d\vec q_1} 
\int\! {d\vec q_2} 
\int\! {d\vec q_3}\  
\delta\!\left(\sum \vec q_i\right)\
\int_0^{M_{\rm soft}} { d|\vec l\,| \over |\vec l\,|}\,
\int { d^2\hat l \over 4 \pi}\
\nonumber\\
&&\times
{\alpha_s \over \pi}\,
F_{ij}^J(\hat l;\hat q_1,\hat q_2, \hat q_3)\,
R_0(\vec q_1,\vec q_2, \vec q_3),
\label{softvirtualmod}
\end{eqnarray}
with $J = L$ or $J = R$ depending on whether the virtual loop was to the
left of the cut or to the right. We do this for each potentially soft
gluon in the virtual loop.

This procedure takes care of the terms in  ${\cal I}[{\rm soft,
virtual}]$ that we needed to subtract from ${\cal I}$. Additionally, 
the singularity of the counterterms matches the singularity of the graph
when either of the gluons become soft (see \cite{beowulfPRD}) so that the
graph with these counterterms has only integrable soft-gluon
singularities.

There are some technical issues to attend to. First, one should match the
integration variables in the loop graph to the integration variables in
the counterterms, so that $\{\vec q_1,\vec q_2,\vec q_3\}$ are the
momenta of the final state partons and $\vec l$ is the momentum of the
potentially soft gluon in each case. Second, one needs to deform the
integration contour for the counterterms according to the
$i\epsilon$ prescriptions indicated. This deformation should closely
match that of the next-to-leading order graph in the limit of small $\vec
l$, as discussed in the Appendix.
 
\section{Secondary showering from NLO graphs}

We have started with the order $\alpha_s^{B+1}$ graphs simply as
specified by the Feynman rules and then, because there are order
$\alpha_s^{B+1}$ effects incorporated in the splittings of the
partons emerging from the $\alpha_s^{B}$ graphs and in the soft gluon
radiation from the $\alpha_s^{B}$ graphs, we have subtracted certain
quantities from the $\alpha_s^{B+1}$ graphs. The result is that the 
subtracted $\alpha_s^{B+1}$ graphs with and without virtual loops are
separately free of infrared divergences.

One can attach a shower to each final state parton
emerging from a subtracted order $\alpha_s^{B+1}$ graph. Each shower
affects the expectation value of the observable being calculated.
However, according to the argument in Ref.~\cite{I}, as long as the
observable is infrared safe, the effect is to add terms proportional to
$\alpha_s^{B+2}$ and higher powers of $\alpha_s$. This holds
independently of the exact functions used to generate the showers, as
long as these functions have the required generic properties. For our
present purposes, we can use the simple algorithm described in Sec.~VIII
of Ref.~\cite{I}.

The only thing that we need to specify is the
choice for width parameter $\kappa$ for the first splitting in the shower.
We will take
\begin{equation}
\kappa^2 = \vec q^{\,2}\ (1-t),
\end{equation}
where $\vec q$ is the momentum of the parton that is to split and $t$ is
the thrust of the partonic final state in the order $\alpha_s^{B+1}$
graph. The factor $(1-t)$ is important. In the case that we generate an
event with thrust very near to 1, we do not want the showering to turn
this two-jet event into a three-jet event, an event with thrust
substantially less than 1. The factor $(1-t)$ prevents this. [For thrust
greater than a value $t_{\rm max}$ close to 1, the code \cite{beowulfcode}
provides a stronger cutoff:
$\kappa^2 = \vec q^{\,2}\ (1-t)^2/(1-t_{\rm max})^2$. The default value
of $t_{\rm max}$ is 0.95.]

\section{Exponentiation}

We now return to the terms that we have added. Consider the sum
\begin{equation}
{\cal I}[{\rm Born}]
+{\cal I}[{\rm soft,real}]+{\cal I}[{\rm soft,virtual}]
\end{equation}
from Eqs.~(\ref{softborn}), (\ref{softreal}), and (\ref{softvirtual}).
Compare this to the quantity
\begin{eqnarray}
{\cal I}[{\rm radiate}] &=&
\int\! {d\vec q_1} 
\int\! {d\vec q_2} 
\int\! {d\vec q_3}\ 
\delta\!\left(\sum \vec q_i\right)\
\int_0^{M_{\rm soft}} { d|\vec l\,| \over |\vec l\,|}\,
\int { d^2\hat l \over 4 \pi}\
\nonumber\\
&&\times
\sum_{ij}
{\alpha_s \over \pi}\,
F_{ij}(\hat l;\hat q_1,\hat q_2, \hat q_3)\,
R_{ij}(\vec l;\vec q_1,\vec q_2, \vec q_3)
\nonumber\\
&&\times
\exp\!\left(-
\int_{|\vec l\,|}^{M_{\rm soft}} { d|\vec l'\,| \over |\vec l'\,|}\,
\int { d^2\hat l' \over 4 \pi}\
\sum_{i'j'}
{\alpha_s \over \pi}\,
F_{i'j'}(\hat l';\hat q_1,\hat q_2, \hat q_3)
\right).
\label{radiate}
\end{eqnarray}
Here the three outgoing partons radiate a soft gluon according to the
radiation pattern specified by the $F_{ij}$. The gluon radiation at small
$\vec l$ is suppressed by the exponential function, which gives the
probability that the gluon was not already radiated with a higher energy
than $|\vec l\,|$. As we found in the companion paper \cite{I} on
collinear emissions, there is a subtraction to multiplication theorem
that says that
\begin{equation}
{\cal I}[{\rm Born}]
+{\cal I}[{\rm soft,real}]+{\cal I}[{\rm soft,virtual}]
= {\cal I}[{\rm radiate}]
\times \left(1 + {\cal O}(\alpha_s^2)\right).
\end{equation}
The proof of this is simple. We have
\begin{equation}
{\cal I}[{\rm radiate}]
={\cal I}[{\rm radiate},1] + {\cal I}[{\rm radiate},2],
\end{equation}
where 
\begin{eqnarray}
{\cal I}[{\rm radiate},1] &=&
\int\! {d\vec q_1} 
\int\! {d\vec q_2} 
\int\! {d\vec q_3}\ 
\delta\!\left(\sum \vec q_i\right)\
\int_0^{M_{\rm soft}} { d|\vec l\,| \over |\vec l\,|}\,
\int { d^2\hat l \over 4 \pi}\
\nonumber\\
&&\times
\sum_{ij}
{\alpha_s \over \pi}\,
F_{ij}(\hat l;\hat q_1,\hat q_2, \hat q_3)\,
[R_{ij}(\vec l;\vec q_1,\vec q_2, \vec q_3)
-R_{0}(\vec q_1,\vec q_2, \vec q_3)]
\nonumber\\
&&\times
\exp\!\left(-
\int_{|\vec l\,|}^{M_{\rm soft}} { d|\vec l'\,| \over |\vec l'\,|}\,
\int { d^2\hat l' \over 4 \pi}\
\sum_{i'j'}
{\alpha_s \over \pi}\,
F_{i'j'}(\hat l';\hat q_1,\hat q_2, \hat q_3)
\right)
\label{radiate1}
\end{eqnarray}
and
\begin{eqnarray}
{\cal I}[{\rm radiate},2] &=&
\int\! {d\vec q_1} 
\int\! {d\vec q_2} 
\int\! {d\vec q_3}\ 
\delta\!\left(\sum \vec q_i\right)\
R_{0}(\vec q_1,\vec q_2, \vec q_3)
\nonumber\\
&&\times
\int_0^{M_{\rm soft}} { d|\vec l\,| \over |\vec l\,|}\,
\int { d^2\hat l \over 4 \pi}\
\sum_{ij}
{\alpha_s \over \pi}\,
F_{ij}(\hat l;\hat q_1,\hat q_2, \hat q_3)\,
\\
&&\times
\exp\!\left(-
\int_{|\vec l\,|}^{M_{\rm soft}} { d|\vec l'\,| \over |\vec l'\,|}\,
\int { d^2\hat l' \over 4 \pi}\
\sum_{i'j'}
{\alpha_s \over \pi}\,
F_{i'j'}(\hat l';\hat q_1,\hat q_2, \hat q_3)
\right).
\nonumber
\label{radiate2}
\end{eqnarray}
Now ${\cal I}[{\rm radiate},1]$ can be expanded in powers of $\alpha_s$
to give
\begin{eqnarray}
{\cal I}[{\rm radiate},1] &=&
\int\! {d\vec q_1} 
\int\! {d\vec q_2} 
\int\! {d\vec q_3}\ 
\delta\!\left(\sum \vec q_i\right)\
\int_0^{M_{\rm soft}} { d|\vec l\,| \over |\vec l\,|}\,
\int { d^2\hat l \over 4 \pi}\
\nonumber\\
&&\times
\sum_{ij}
{\alpha_s \over \pi}\,
F_{ij}(\hat l;\hat q_1,\hat q_2, \hat q_3)\,
[R_{ij}(\vec l;\vec q_1,\vec q_2, \vec q_3)
-R_{0}(\vec q_1,\vec q_2, \vec q_3)]
\nonumber\\
&&+
{\cal O}(\alpha_s^2 \times R).
\label{radiate1new}
\end{eqnarray}
This is just ${\cal I}[{\rm soft,real}]+{\cal I}[{\rm soft,virtual}]$.
The integral in ${\cal I}[{\rm radiate},2]$ looks complicated, but it is
simple because the $|\vec l\,|$ integral is the integral of a derivative.
Thus ${\cal I}[{\rm radiate},2]$ is the difference of the integrand
between the integration end points:
\begin{equation}
{\cal I}[{\rm radiate},2] =
\int\! {d\vec q_1} 
\int\! {d\vec q_2} 
\int\! {d\vec q_3}\ 
\delta\!\left(\sum \vec q_i\right)\
R_{0}(\vec q_1,\vec q_2, \vec q_3)
[F(M_{\rm soft}) - F(0)],
\end{equation}
where
\begin{eqnarray}
F(|\vec l\,|) &=& \exp\!\left(-
\int_{|\vec l\,|}^{M_{\rm soft}} { d|\vec l'\,| \over |\vec l'\,|}\,
\int { d^2\hat l' \over 4 \pi}\
\sum_{i'j'}
{\alpha_s \over \pi}\,
F_{i'j'}(\hat l';\hat q_1,\hat q_2, \hat q_3)
\right)
\nonumber\\
&=&\exp\left(- (\alpha_s/\pi)\, C \log(M_{\rm soft}/|\vec l\,|)\right).
\end{eqnarray}
Here $C$ is the angular integral in the exponent, given in 
Eq.~(\ref{angularint}). Now $F(M_{\rm soft}) = \exp(-0) = 1$. On the
other hand, the exponent diverges as $|\vec l\,| \to 0$. Furthermore, $C$
is positive. Thus $F(|\vec l\,|) \to \exp(- \infty) = 0$ as $|\vec l\,|
\to 0$. The result for ${\cal I}[{\rm radiate},2]$ is then
\begin{equation}
{\cal I}[{\rm radiate},2] =
\int\! {d\vec q_1} 
\int\! {d\vec q_2} 
\int\! {d\vec q_3}\ 
\delta\!\left(\sum \vec q_i\right)\
R_{0}(\vec q_1,\vec q_2, \vec q_3).
\label{radiate2new}
\end{equation}
This completes the proof of the theorem.

\section{The result}

We are now able to generate a useful expression for the observable
${\cal I}$. We begin with a cut graph that contributes to ${\cal I}[{\rm
Born}]$, Eq.~(\ref{softbornfull}). For each of the final state partons,
we introduce an additional integration,
\begin{equation}
\int_0^{\lambda_V \vec q_i^{\,2}} {d\bar q_i^2 \over \bar q_i^2}
\int_0^1\! dx_i
\int_{-\pi}^\pi\! {d\phi_i\over 2\pi}\
\frac{\alpha_s}{2\pi}{\cal M}_{i}(\bar q_i^2,x_i,\phi_i)\
\exp\!\left(-\int_{\bar q_i^2}^\infty{ d\bar l_i^2 \over \bar l_i^2}
\int_0^1\! dz_i\
{ \alpha_s \over 2\pi}\,{\cal P}_{\!i}(\bar l_i^2, z_i)
\right),
\end{equation}
representing the splitting of that parton, as explained in Ref.~\cite{I}.
We also introduce an integration,
\begin{eqnarray}
&&
\int_0^{M_{\rm soft}} { d|\vec l\,| \over |\vec l\,|}\,
\int { d^2\hat l \over 4 \pi}\
\sum_{ij}
{\alpha_s \over \pi}\,
F_{ij}(\hat l;\hat q_1,\hat q_2, \hat q_3)\,
\nonumber\\
&&\times
\exp\!\left(-
\int_{|\vec l\,|}^{M_{\rm soft}} { d|\vec l'\,| \over |\vec l'\,|}\,
\int { d^2\hat l' \over 4 \pi}\
\sum_{i'j'}
{\alpha_s \over \pi}\,
F_{i'j'}(\hat l';\hat q_1,\hat q_2, \hat q_3)
\right),
\end{eqnarray}
representing the radiation of a soft gluon from the final state partons.
The function representing the Born matrix element then depends on all of
the integration variables. This gives a probability to create a total of
seven partons. Each of these seven partons then serves as the source of a
secondary shower.\footnote{For the soft gluon, we choose the starting
scale for the secondary shower to be $\kappa^2 = c_{\kappa,{\rm
soft}}\,\vec l^{\,2}$ where $\vec l$ is the momentum of the soft gluon and
$c_{\kappa,{\rm soft}}$ is a constant taken to be 1/10 in the numerical
example in the following section.} This is illustrated in
Fig.~\ref{fig:shower2}.

In addition, there are contributions from order $\alpha_s^{B+1}$ cut
Feynman graphs plus their soft gluon counterterms. Included here are cut
self-energy graphs, but only for virtuality above the cutoff $\lambda_V
\vec q_i^{\, 2}$, so that there are no collinear divergences. Each of the
three of four partons emerging from such a graph then serves as the
source of a secondary shower. These order $\alpha_s^{B+1}$ remnant terms
are complicated, but they have only integrable singularities and they are
just what is needed to leave the calculation correct to next-to-leading
order.

\section{A numerical test}

In this section I present some numerical results based on the algorithm
described above. I have computed the thrust distribution
$d\sigma/dt$ for thrust $t = 0.86$, in the middle of the
three-jet region. We will compare this distribution calculated
with parton showers to the same distribution calculated with a
straightforward NLO computation with no showers. The difference should be
of order $\alpha_s^{B+2}$. We divide the difference by the NLO result,
forming
\begin{equation}
R = \frac{\mbox{\rm (NLO-shower)} - {\rm NLO}}{{\rm NLO}}.
\label{Rdef}
\end{equation}
The ratio $R$ should have a perturbative expansion that begins at order
$\alpha_s^2$. We can test this by plotting the ratio against 
$\alpha_s^2$. We expect to see a curve that approximates a straight line
through zero for small $\alpha_s^2$. For comparison, I exhibit also the
ratio with the order $\alpha_s^B$ graphs with showers but with the
$\alpha_s^{B+1}$ corrections omitted,
\begin{equation}
R_{\rm LO} = \frac{\mbox{\rm (LO-shower)} - {\rm NLO}}{{\rm NLO}}.
\label{RLOdef}
\end{equation}
The ratio $R_{\rm LO}$ should have a perturbative expansion that
begins at order $\alpha_s^1$. Thus we expect to see a curve proportional
to the square root function $[{\alpha_s^2}]^{1/2}$ for small $\alpha_s^2$.
The size of the coefficient of $[{\alpha_s^2}]^{1/2}$ in $R_{\rm LO}$
does not have any great significance, since it is quite sensitive to the
choice of renormalization scale.

\begin{figure}
\includegraphics[width = 8 cm]{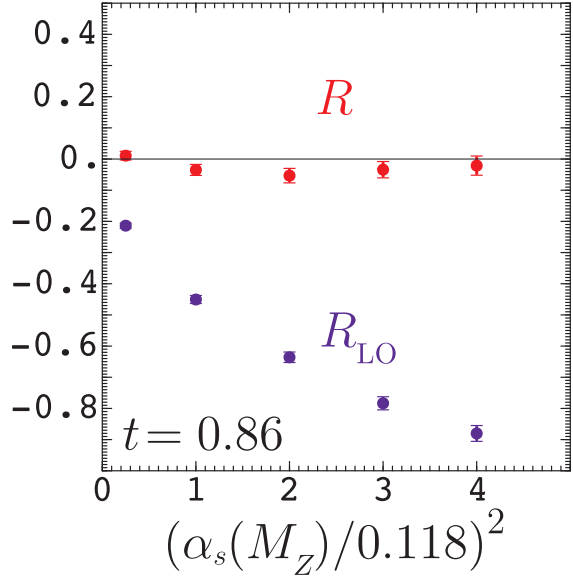}
\caption{Comparison of the NLO calculation with showers added as
described in this paper and \protect\cite{I} to a pure NLO calculation
using \protect\cite{beowulfcode}.  I plot the ratio $R$ defined in
Eq.~(\ref{Rdef}) for the thrust distribution at thrust equal 0.86. Also
shown is the ratio  $R_{\rm LO}$, defined in Eq.~(\ref{RLOdef}), in which
the order $\alpha_s^{B+1}$ correction terms are omitted from the
calculation. The c.m.\ energy is $\sqrt S = M_Z$ and the renormalization
scale is chosen to be $\mu = \sqrt S/6$. These ratios are calculated for
$\alpha_s(M_Z)^2 = \{0.25,1,2,3,4\}\times (0.118)^2$ and plotted versus
$\alpha_s(M_Z)^2/(0.118)^2$.}
\label{fig:test}
\end{figure}

The results of this test are shown in Fig.~\ref{fig:test} for $\sqrt S =
M_Z$ with a renormalization scale $\mu = \sqrt S/6$. The values of
$\alpha_s(M_Z)$ range from  $(1/2)\times 0.118$ to $2\times 0.118$, where
0.118 represents something close to the physical value for
$\alpha_s(M_Z)$. One should note that $2\times 0.118$ amounts to quite a
large $\alpha_s$ in this calculation since $\alpha_s(M_Z) = 0.236$ gives
$\alpha_s(\mu) = 0.586$.

We see the expected shape of the $R_{\rm LO}$ curve. We then examine
whether the $R$ curve approaches a straight line through the origin as
$\alpha_s^2 \to 0$. Within the errors, it does. However, the slope of the
straight line is quite small. Presumably with other choices of parameters
in the program, the absolute value of the slope would be bigger.

\begin{figure}
\includegraphics[width = 8 cm]{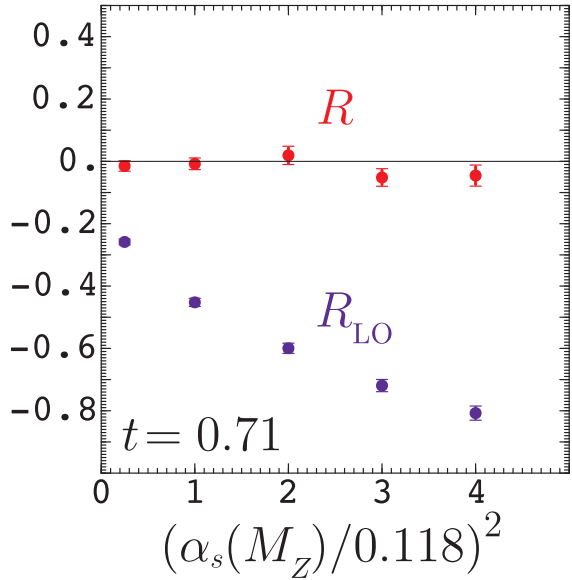}
\caption{Comparison of the NLO calculation with showers added to
a pure NLO calculation for $t = 0.71$. The notation is as in
Fig.~\ref{fig:test}.}
\label{fig:test71}
\end{figure}

In Fig.~\ref{fig:test71}, I show the same comparison, but this time for $t
= 0.71$. This is near the value $t = 2/3$ that marks the far end of the
three-jet region, with the three partons in what is sometimes called
the Mercedes configuration. The results are similar to the results for $
t = 0.86$.

\begin{figure}
\includegraphics[width = 8 cm]{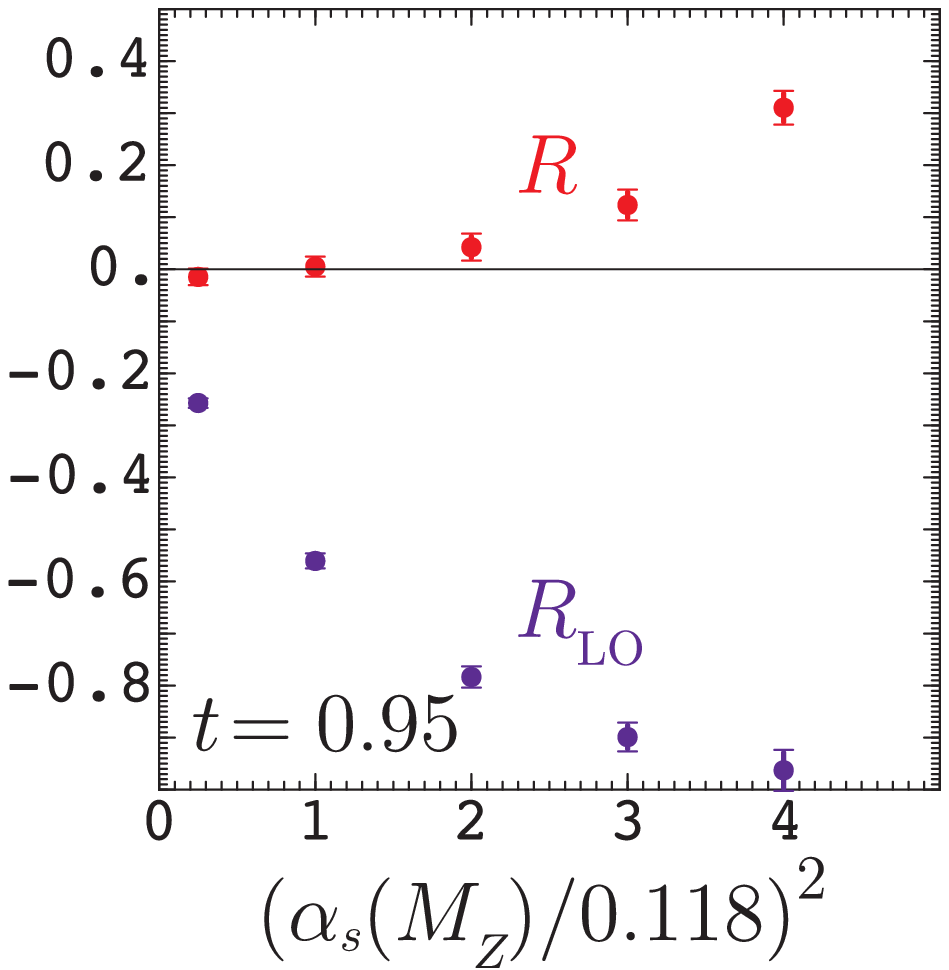}
\caption{Comparison of the NLO calculation with showers added to
a pure NLO calculation for $t = 0.95$. The notation is as in
Fig.~\ref{fig:test}.}
\label{fig:test95}
\end{figure}

In Fig.~\ref{fig:test95}, I show the same comparison, but this time for
$t = 0.95$. This is near the two-jet limit at $t = 1$. For $\alpha_s =
0.118$, one would normally not use a calculation that did not include a
summation of logs of $1-t$ for $t$ this close to 1, since $\log(0.05)^2
\approx 9$. Thus I would not recommend using the code discussed in this
paper for a comparison to data this near to the two-jet limit.
Nevertheless, we can still test for the absence of an $\alpha_s^{1}$ term
in $R$. Looking at the graph, we see that, within the errors, there is no
evidence for a nonzero $\alpha_s^{1}$ term in $R$. We expect an
$\alpha_s^{2}$ term, but it appears that the coefficient of $\alpha_s^{2}$
is quite small.  On the other hand, it appears that some of the yet
higher order terms are quite substantial.

\section{Areas where more work is needed}

The present work suffers from a number of deficiencies. First, the code is
designed to calculate three-jet cross sections in electron-positron
annihilation correctly to NLO, but it ignores the two-jet cross section.
It should be rather simple to calculate the two-jet cross section at
next-to-leading order with showers. Then one should merge the two
calculations so that three-jet observables are calculated correctly at
order $\alpha_s^2$ while the two-jet cross section is calculated
correctly at order $\alpha_s^1$. 

Such a merged calculation would have benefits even for the calculation of
three-jet observables. Specifically, there would be an advantage when
calculating three-jet observables close to the two-jet region. For
example, consider the thrust distribution near to $t = 1$. The two-jet
calculation with showers, before any merging, would approximately sum
terms proportional to $\alpha_s^N\,\log^J(1-t)$, generating these terms by
showering from the two-jet hard process. For $d\sigma / dt$, the merged
calculation should contain the $\alpha_s^1$ and $\alpha_s^2$ contributions
correctly, including of course the term proportional to $\alpha_s^2\,
\log(1-t)$. It should also include the terms $\alpha_s^n \times{\rm
logs}$ for $n>2$ generated by showering from a two-jet configuration. The
three-jet algorithm presented in this paper does not include merging with
a two-jet calculation and thus does not generate these logarithms. Thus
the present program, when applied to the thrust distribution, should not
be used for small values of $1-t$.

Second, it is desirable to add not only
parton showers but hadronization to NLO perturbative calculations. For
hadronization, one can use an existing model \cite{Pythia,Herwig,Ariadne}.
However, most hadronization models require color information on the final
state partons. Since the current code does not assign a color structure to
these partons, it will be necessary to add this capability. 

Third, in
principle, hadronization models should be infrared safe in the sense that
final partonic states that differ by a parton splitting into two almost
collinear partons (or one parton and another with almost zero
momentum) will produce almost the same hadronic final state. Since
hadronization models currently in use do not necessarily have this
property, the NLO calculation should provide it by combining parton pairs
with virtuality less than some cutoff parameter matched to the
hadronization model to be used. 

Finally, the current paper includes
hardly any numerical investigations to check how well the program works
for various observables and how the results depend on parameters such as
the renormalization scale $\mu$ and the soft gluon limiting energy
$M_{\rm soft}$. Addressing these needs remains a topic for future
research.

\section{Conclusions}

In this paper and \cite{I}, we have seen how to add parton showers to a
next-to-leading order calculation in QCD in such a way that the result
obtained for an infrared safe observable remains correct to
next-to-leading order. The main feature of this procedure is to turn the
collinear singularities of the order $\alpha_s^{B+1}$ graphs into the
primary parton splittings, that is the first splittings of the partons
that emerge from an $\alpha_s^{B}$ graph. This was accomplished in
\cite{I} in a particularly simple fashion by virtue of the fact that
Ref.~\cite{I} is based on a NLO calculation in the Coulomb gauge. In this
gauge the collinear and collinear $\times$ soft singularities are
entirely contained in the cut self-energy graphs. Thus we simply
eliminate the small-virtuality part of cut self-energy subgraphs in
$\alpha_s^{B+1}$ graphs and instead incorporate these contributions into
the primary parton splittings from the corresponding  $\alpha_s^{B}$
graph. One could subtract a different quantity from the $\alpha_s^{B+1}$
graphs as long as it has the same collinear and collinear $\times$  soft
singularities,\footnote{This might, however, require adjustment of the
treatment of the soft gluon singularities presented in this paper.} as in
the work of Refs.~\cite{FrixioneWebberI, FrixioneWebberII}.

The next step, and the main subject of this paper, is to treat the soft,
wide angle singularities of the theory. Again, we subtract terms with
these singularities from the $\alpha_s^{B+1}$ graphs and then incorporate
them as radiation from the $\alpha_s^{B}$ graphs. Here, the hard partons
emerging from an $\alpha_s^{B}$ scattering form an antenna that radiates
the soft gluon. The gluon is radiated coherently from the antenna and is
not part of any jet.

This gives an algorithm that can be viewed as a simulation of the
production of three hard partons (for the three-jet process considered in
this paper) and one soft gluon, with each of the three hard partons
splitting into two daughter partons. The six daughter partons and one
soft gluon then serve as the progenitors of parton showers according to
the simple algorithm used in Ref.~\cite{I} or any standard algorithm. 
There remains the remnants of the order $\alpha_s^{B+1}$ graphs, with
three and four parton final states. Each of these final state partons
serves as the progenitor of whatever sort of parton shower is desired.

The results of this paper suggest a strategy for building future NLO
calculations matched to parton showers. The matching concerns the primary
parton splittings, that is the first splittings of the partons emerging
from a Born graph. The matching also needs the radiation of a soft gluon
from the antenna formed by partons emerging from the Born graph. This
suggests that the author of an NLO calculation might build in the primary
parton splittings and the soft gluon radiation. There are also secondary
splittings. These are the splittings from the soft gluon produced from a
Born graph, the splittings of the daughters produced by the primary
splittings, the splittings of the partons from an order
$\alpha_s^{B+1}$ graph, and, finally, the further splittings of all of
these partons as a complete parton shower is formed. The secondary
splittings do not need to be matched to the NLO calculation. That is, the
soft gluon and the partons after the primary splittings and the partons
produced by the $\alpha_s^{B+1}$ graphs can be fed to any shower and
hadronization algorithm as long as it is infrared safe. The result is
that for a given process such as $e^+ + e^- \to 3\ {\rm jets}$ we can
have $N_{\rm NLO}$ next-to-leading order calculations and $N_{\rm MC}$
shower and hadronization calculations and we do not need $N_{\rm
NLO}\times N_{\rm MC}$ matching schemes as long as each NLO calculation
includes its primary parton splittings and treatment of soft gluon
radiation.

This does not preclude debate over how best to do the primary parton
splitting and soft gluon radiation. This paper and Ref.~\cite{I} have, of
necessity, chosen one scheme, but that scheme may well be less than
optimal.

The code described in this paper is available at \cite{beowulfcode}. 

\acknowledgements

The author is pleased to thank his collaborator on Ref.~\cite{I}, Michael
Kr\"amer, for his advice. In addition, Steve Mrenna, John Collins,
Torbj\"orn Sj\"ostrand, George Sterman, and Michael Seymour provided
helpful advice and criticisms. This work was supported in part by the
U.S.~Department of Energy, by the European Union under contract
HPRN-CT-2000-00149, and by the British Particle Physics and Astronomy
Research Council.

\appendix*

\section{The integration for virtual soft gluons}

We have seen that in order to calculate the contribution to ${\cal I}$
from an order $\alpha_s^2$ graph with a virtual loop, we should calculate
a difference integral of the form
\begin{eqnarray}
&&
\int\! {d\vec q_1} 
\int\! {d\vec q_2} 
\int\! {d\vec q_3}\  
\delta\!\left(\sum \vec q_k\right)\
\biggl\{\int  d^3\vec l\
f(\vec l,\vec q_1,\vec q_2, \vec q_3)
\nonumber\\
&&-
\sum_{\{i,j\}}\int  {d^3\vec l' \over  2|\vec l'|^3}\,
\theta(l^{\prime 2} < M_{\rm soft}^2)
{\alpha_s \over 2\pi^2}\,
F_{ij}^J(\hat l';\hat q_1,\hat q_2, \hat q_3)\,
R_{0}(\vec q_1,\vec q_2, \vec q_3)
\biggr\},
\label{softsubtraction}
\end{eqnarray}
where $J$ is $L$ or $R$ and $F_{ij}^L$ and $F_{ij}^R$ are given in
Eqs.~(\ref{FijL}) and (\ref{FijR}). In this appendix, we consider $J=L$,
that is a virtual loop to the left of the final state cut. The integrand
for the original graph is represented here by the function $f$. The
momenta $\vec q_i$ are the momenta of the final state partons, while
$\vec l$ is the momentum in the virtual loop. There are either one or two
terms in the sum over the indices $\{i,j\}$, depending on the graph. The
term $\{i,j\}$ is included if there is a potentially soft virtual gluon
connecting final state lines $i$ and $j$ in $f$. Let $\vec l_{ij}$ be that
combination of $\vec l$ and the $\vec q_k$ that is carried by this
potentially soft virtual line in $f$. For example if the final state
propagators 1 and 2 are connected by a gluon line that carries momentum
$l$ in the virtual loop and propagators 2 and 3 are also connected by a
virtual gluon line that carries momentum $l - q_2$, then  $\vec l_{12} =
\vec l$ and $\vec l_{23} = \vec l - \vec q_2$. In the virtual subtraction
term, we call the integration variable $\vec l'$. If we identify $\vec
l'$ with $\vec l_{ij}$, the $F_{ij}$ subtraction term matches $f$ when
$\vec l_{ij} \to 0$, so that the leading singularity is cancelled. 

The method used in this paper is to perform the virtual loop integrations
numerically, as in \cite{KSCoulomb,beowulfPRL,beowulfPRD,beowulfrho} and
\cite{I}. Thus we must make sure that this cancellation works
numerically. First, we define the dependence of $f$ and $R$ on $\vec q_1,
\vec q_2, \vec q_3$ to be the dependence computed from the relevant
Feynman graphs with the final state momenta $q_1, q_2, q_3$ on-shell:
$q_k^2 = 0$. Of course, in our present discussion, the $q_k$ are the
momenta of on-shell massless partons, so evidently $q_k^2 = 0$. However,
at a later stage in the calculation each of these partons will generate a
parton shower with the same $\vec q$. Even with a shower, we take the
functions $f$ and $R$ to remain unchanged. This is in contrast to our
treatment of showering from the Born graphs and is required for the NLO
graphs in order to preserve the soft gluon cancellations.

Next, in the integral of the full graph $f$, the integration is
deformed\footnote{Some relevant theorems concerning contour deformations
in more than one dimension are given in \cite{beowulfPRD}.} into complex
loop-momentum space,
\begin{equation}
\int  d^3\vec l\ {\cal J}(\vec l\,)\
f\!\left(\vec l + i\vec \kappa(\vec l\,),\vec q_1,\vec q_2, \vec
q_3\right).
\end{equation}
The imaginary part of the deformed loop momentum is a definite function
$\vec \kappa(\vec l)$ of the real part. The Jacobian ${\cal J}$ is the
determinant of the matrix $(\delta_{IJ} + i \partial \kappa^I/\partial
l^J)$. By deforming the contour, we avoid the singularity of the form
\begin{equation}
{ 1 \over |\vec q_i| + |\vec q_j| - 
|\vec q_i + \vec l_{ij} + i\vec\kappa| 
- |\vec q_j - \vec l_{ij} - i\vec\kappa|
+ i\epsilon}
\label{softsing}
\end{equation}
in $f$, the $\{i,j\}$ {\it scattering singularity}. It is, however, not
allowed to deform the contour away from the soft singularity at $\vec
l_{ij} = 0$. For that reason $\vec \kappa(\vec l\,)$ is proportional to
$\vec l_{ij}^2$  as $\vec l_{ij}^2 \to 0$. Hence we come very near to the
scattering singularity when $\vec l_{ij}$ is small.

What about the subtraction term? Here there is a singularity of the form
\begin{equation}
{ 1 \over \vec l' \cdot (\hat q_j - \hat q_i) + i \epsilon}.
\label{approxsoftsing}
\end{equation}
We may call this the {\it approximate scattering singularity}. We note
that
\begin{equation}
|\vec q_i| + |\vec q_j| - 
|\vec q_i + \vec l_{ij} | - |\vec q_j - \vec l_{ij}|
\sim \vec l_{ij} \cdot (\hat q_j - \hat q_i)
\label{softsingapprox}
\end{equation}
for $\vec l_{ij} \to 0$, so the singularity (\ref{approxsoftsing}) is
indeed an approximation to Eq.~(\ref{softsing}) if we identify $\vec l'$
with
$\vec l_{ij}$. As in the $f$ integral, we need to deform the contour to
avoid the singularity (\ref{approxsoftsing}). As in the $f$ integral, the
deformation will have to vanish as we approach the soft singularity at
$\vec l' = 0$. In order for the cancellation between the full graph and
the soft gluon subtraction to work, we need to be careful about the
matching of the singularities in the two integrands.

We are now in a position to state the matching problem more precisely. We
define complex vectors $s_{ij}$, one for each potentially soft gluon line
$\{i,j\}$ in Eq.~(\ref{softsubtraction}). Each $s_{ij}$ is a function of
the corresponding $l_{ij}$ that is asymptotically equal to $l_{ij}$ when
$l_{ij}$ is small. Then we write the integral in 
Eq.~(\ref{softsubtraction}) in the form
\begin{eqnarray}
&&
\int\! {d\vec q_1} 
\int\! {d\vec q_2} 
\int\! {d\vec q_3}\  
\delta\!\left(\sum \vec q_i\right)\
\int  d^3\vec l\ \biggl\{
{\cal J}(\vec l\,)\
f(\vec l + i\vec\kappa,\vec q_1,\vec q_2, \vec q_3)
\nonumber\\
&&-
\sum_{ij} {{\cal J}_{\!s}(\vec l_{ij}) \over  2|\vec s_{ij}|^3}\,
\theta(\vec l_{ij}^{\, 2} < M_{\rm soft}^2)\,
{\alpha_s \over 2\pi^2}\,
F_{ij}^J(\hat s_{ij};\hat q_1,\hat q_2, \hat q_3)\,
R_{0}(\vec q_1,\vec q_2, \vec q_3)
\biggr\}.
\label{softsubtraction2}
\end{eqnarray}
The Jacobian ${\cal J}_s$ is the determinant of the matrix $\partial
s_{ij}^I/\partial l_{ij}^J$.  Our goal will be to specify the functions
$s_{ij}$ so that we avoid the approximate scattering singularity in the
second term of Eq.~(\ref{softsubtraction2}) and so that the $\vec l_{ij}
\to 0$ singularities in the sum of the two terms cancel sufficiently well
that the integral is convergent.

We will approach this problem in stages. First, we concentrate on the 
$\vec l_{ij} \to 0$ region.

We immediately recognize a small problem. The soft singularity (with $\vec
\kappa = 0$) lies on an ellipse, while the approximate soft singularity
lies on a plane. To make them match at the required accuracy, we define
\begin{equation}
\vec s_{ij} = \vec l_{ij} 
+ 
(1 - \vec l_{ij}^{\,2}/M_{\rm soft}^2)
\left\{\vec\xi(\vec l_{ij}) + i \vec \zeta(\vec l_{ij})
\right\}
\hskip 1 cm{\rm (to\ be\ improved)},
\label{sijstart}
\end{equation}
where $\vec\xi$ and $\vec \zeta$ are certain functions of $\vec
l_{ij}$ to be defined below and $M_{\rm soft}^2$ is the maximum value of
$\vec l^{\,2}$ in the soft gluon subtraction, Eq.~(\ref{softrealfull}).

The idea is to have the soft subtraction cancel the leading singular
behavior of $f$ point by point in $\vec l$. To do this, we need to match
the singularity
\begin{equation}
{ 1 \over D_0(\vec l_{ij} + i\vec\kappa)} =
{ 1 \over |\vec q_i| + |\vec q_j| - 
|\vec q_i + \vec l_{ij} +i\vec\kappa| 
- |\vec q_j - \vec l_{ij}+ i\vec\kappa|}
\end{equation}
in $f$ with the singularity
\begin{equation}
{ 1 \over D_s(\vec s_{ij})} =
{ 1 \over \vec  s_{ij} \cdot (\hat q_j - \hat q_i)}
\end{equation}
in $F_{ij}$.

There are three parts in the relation (\ref{sijstart}) between $\vec
s_{ij}$ and $\vec l_{ij}$. There is a real displacement $\vec \xi$, a
contour deformation $\vec \zeta$ in the imaginary direction, and a factor
$(1 - \vec l_{ij}^{\,2}/M_{\rm soft}^2)$. The factor $(1 - \vec
l_{ij}^{\,2}/M_{\rm soft}^2)$ serves to set the deformation to zero at
the edge of the integration region and needs no further discussion. We
choose the real displacement to be
\begin{equation}
\vec \xi(\vec l_{ij}) = 
{ \vec l_{ij}^{\,2} \over 
2 |\vec q_i| |\vec q_j|[1 - \hat q_i \cdot \hat q_j]}
\left\{
[1 - (\hat l_{ij}\cdot \hat q_j)^2]\, \vec q_i
-
[1 - (\hat l_{ij}\cdot \hat q_i)^2]\, \vec q_j
\right\}.
\label{xidef}
\end{equation}
This choice is motivated by carrying out the expansion
(\ref{approxsoftsing}) to one more order. Then one finds that
\begin{equation}
(\vec l_{ij} + \vec \xi (\vec l_{ij}))\cdot (\hat q_j - \hat q_i)
\sim
|\vec q_i| + |\vec q_j| - 
|\vec q_i + \vec l_{ij}| - |\vec q_j - \vec l_{ij}|
+{\cal O}(l_{ij}^3).
\end{equation}
The imaginary displacement is
\begin{equation}
\vec \zeta(\vec l\,) = 
C_\zeta\,\vec l^{\,2}
\min\!\left((1 + \hat l\cdot  \hat q_i),(1 - \hat l\cdot  \hat q_j)\right)
(\hat q_j - \hat q_i).
\label{zetadef}
\end{equation}
Here $C_\zeta$ is a parameter with dimensions of inverse momentum. This
function matches the function $\vec \kappa(\vec l\,)$ in the limit of
small $\vec l$ as long as one chooses $C_\zeta$ to be the same parameter
as is used in $\vec \kappa(\vec l\,)$.\footnote{Specifically, $C_\zeta =
2\alpha/[(1+\gamma)(|\vec q_i| + |\vec q_j| - |\vec q_i+\vec q_j|)]$,
where $\alpha$ and $\gamma$ are the parameters used to define the
deformation in \cite{{beowulfPRD}}.} Thus
\begin{equation}
D_s(\vec s_{ij}) = 
D_0(\vec l_{ij} + i\vec\kappa)
+{\cal O}(l_{ij}^3).
\end{equation}

There are other singularities that have to match. First, both $f$ and the
soft subtraction terms have $1/|\vec l_{ij}|^3$ soft singularities that
evidently match since $\vec s_{ij} = \vec l_{ij}$ to lowest order.
Second, both terms have collinear singularities along the lines $1 + \hat
l_{ij} \cdot \hat q_i = 0$ and $1 - \hat l_{ij} \cdot \hat q_j = 0$.
These are pinch singularities: it is not allowed to deform the contours
to avoid them. For this reason $\vec \zeta(\vec l)$ vanishes when $\vec
l$ is in the same direction as $-\vec q_i$ or $\vec q_j$. Furthermore,
the part of $\vec \xi$ that is not proportional to $\vec q_i$ vanishes
when   $\vec l$ is in the same direction as $-\vec q_i$ and the part of
$\vec \xi$ that is not proportional to $\vec q_j$ vanishes when  
$\vec l$ is in the same direction as $\vec q_j$.

We are now ready for the second stage of our construction. In the first
stage, we concentrated on the $\vec l_{ij} \to 0$ region. The function
$\vec s_{ij}$ given in Eq.~(\ref{sijstart}) works fine for small $\vec
l_{ij}$. However, our integration region ends at $|\vec l_{ij}| = M_{\rm
soft}$, so we supplied a factor that turns the contour deformation off at
the boundary of the integration region.\footnote{This is the analogue of
keeping the contour fixed at the end point of the integration in a one
dimensional contour integration.} Unfortunately, the singularity at 
$\vec  l_{ij} \cdot (\hat q_j - \hat q_i) = 0$ intersects the surface
$|\vec l_{ij}| = M_{\rm soft}$. Thus we should retain some contour
deformation even at the boundary $|\vec l_{ij}| = M_{\rm soft}$ of the
integration region. This is simple to do. 

Let us consider the problem anew, this time ignoring the special problems
at $|\vec l\,| \to 0$ that were addressed in the analysis above. Let
$\vec n = (\hat q_j - \hat q_i)/|\hat q_j - \hat q_i|$ and write $\vec
l_{ij}$ in spherical polar coordinates $(l,\theta,\phi)$ based on $\vec
n$ as the $z$ axis. Then we have an integral
\begin{equation}
\int_0^{M_{\rm soft}}\!l^2\, dl\, \int_{-1}^1\!dx \, \int_{-\pi}^\pi\!
d\phi,
\end{equation}
where $x = \cos\theta$, that is
\begin{equation}
x = \vec l_{ij}\cdot \vec n/|\vec l_{ij}|.
\end{equation}
%.
The integrand has a singularity of the form $1/(x + i\epsilon)$. All that
we need to do is deform the contour for the $x$-integral away from the
singularity. There are collinear singularities at $\hat l_{ij}\cdot \hat
q_j = 1$ and $\hat l_{ij}\cdot \hat q_i = -1$, that is at $\phi = 0,\pi$
and $x = x_0$, where
\begin{equation}
x_0 = \vec n\cdot \hat q_j = \sqrt{(1 - \hat q_i\cdot \hat q_j)/2}.
\end{equation}
The contour cannot be moved at the collinear singularities. Thus we can
take 
\begin{equation}
x \to z = x + i y,
\end{equation}
where
\begin{equation}
y = (\vec l_{ij}^{\,2}/M_{\rm soft}^2)\ (x_0^2 - x^2)^2\, \theta(x^2 <
x_0^2).
\end{equation}
The factor $\vec l_{ij}^{\,2}/M_{\rm soft}^2$ is included here simply to
turn this deformation off for small $\vec l_{ij}^{\,2}$, since we already
have a method for dealing with the small  $\vec l_{ij}^{\,2}$ region. In a
coordinate frame independent notation, this is $\vec l_{ij} \to \vec
s_{ij}$ with
\begin{equation}
\vec s_{ij} = \vec l_{ij} 
+ 
\vec \eta(\vec l_{ij})
\hskip 1 cm{\rm (to\ be\ improved)},
\label{sijsecond}
\end{equation}
where
\begin{equation}
\vec\eta = i y\, |\vec l_{ij}| \vec n
+ \left(
\sqrt{\frac{1-z^2}{1-x^2}} - 1
\right)
\left(\vec l_{ij} - (\vec l_{ij} \cdot \vec n)\,\vec n\right).
\label{etadef}
\end{equation}

Now we have two methods for moving the integration contour, each of which
works in its region and turns off in the region of the other method. All
that we have to do is to add the two deformations,
\begin{equation}
\vec s_{ij} = \vec l_{ij} 
+ 
(1 - \vec l_{ij}^{\,2}/M_{\rm soft}^2)
\left\{\vec\xi(\vec l_{ij}) + i \vec \zeta(\vec l_{ij})
\right\}
+
\vec \eta(\vec l_{ij}).
\label{sijnet}
\end{equation}
This is the final prescription, with $\vec \xi$ given in
Eq.~(\ref{xidef}), $\vec \zeta$ given in Eq.~(\ref{zetadef}), and $\vec
\eta$ given in Eq.~(\ref{etadef}).

%23456789012345678901234567890123456789012345678901234567890123456789012
\end{document}